\documentclass[%
reprint,
superscriptaddress,
 amsmath,amssymb,
 aps,
]{revtex4-1}

\usepackage{graphicx}
\usepackage{dcolumn}
\usepackage{bm}
\usepackage[flushleft]{threeparttable} 

\usepackage{dsfont}
\usepackage{float}
\usepackage{color,psfrag}
\usepackage{mathtools,amscd}

\usepackage{soul}
\usepackage{xcolor} 

\newcommand\here[1]{\fcolorbox{red}{red}{\rule{0pt}{6pt}\rule{6pt}{0pt}}\quad}

\usepackage{ifpdf}
    \ifpdf
\usepackage{tikz}
\usetikzlibrary{arrows,chains,matrix,positioning,scopes, fit, calc}
\usetikzlibrary{cd}
\usepackage{chemfig}
\fi
\usepackage{hyperref}

\hypersetup{
	colorlinks,
	linkcolor=blue,
	citecolor=blue, 
	filecolor=black,
	urlcolor=blue}

\begin{document}

\preprint{APS/123-QED}


\title{Delta--learned force fields for nonbonded interactions: Addressing the strength mismatch between covalent-nonbonded interaction for global models}
\author{Leonardo C\'azares-Trejo}
\affiliation{Instituto de F\'isica, 
Universidad Nacional Aut\'onoma de M\'exico, Cd. de M\'exico C.P. 04510, Mexico}
\author{Marco Loreto-Silva}
\affiliation{Instituto de F\'isica, 
Universidad Nacional Aut\'onoma de M\'exico, Cd. de M\'exico C.P. 04510, Mexico}
\author{Huziel E. Sauceda}
\email{huziel.sauceda@fisica.unam.mx}
\affiliation{Instituto de F\'isica, 
Universidad Nacional Aut\'onoma de M\'exico, Cd. de M\'exico C.P. 04510, Mexico}

\date{\today}

\begin{abstract}
Noncovalent interactions—vdW dispersion, hydrogen/halogen bonding, ion–$\pi$, and $\pi$–stacking—govern structure, dynamics, and emergent phenomena in materials and molecular systems, yet accurately learning them alongside covalent forces remains a core challenge for machine-learned force fields (MLFFs). This challenge is acute for global models that use Coulomb-matrix (CM) descriptors compared under Euclidean/Frobenius metrics in multifragment settings. We show that the mismatch between predominantly covalent force labels and the CM’s overrepresentation of intermolecular features biases single-model training and degrades force-field fidelity. To address this, we introduce \emph{$\Delta$-sGDML}, a scale-aware formulation within the sGDML framework that explicitly decouples intra- and intermolecular physics by training fragment-specific models alongside a dedicated binding model, then composing them at inference. Across benzene dimers, host–guest complexes (C$_{60}$@buckycatcher, NO$_3^-$@i-corona[6]arene), benzene–water, and benzene–Na$^+$, \mbox{$\Delta$-sGDML} delivers consistent gains over a single global model, with fragment-resolved force-error reductions up to \textbf{75\%}, without loss of energy accuracy. Furthermore, molecular-dynamics simulations further confirm that the $\Delta$-model yields a reliable force field for C$_{60}$@buckycatcher, producing stable trajectories across a wide range of temperatures (10-400~K), unlike the single global model, which loses stability above $\sim$200~K. The method offers a practical route to homogenize per-fragment errors and recover reliable noncovalent physics in global MLFFs.
\end{abstract}

\pacs{Valid PACS appear here}
\maketitle



\section{Introduction}

\begin{figure*}[ht]
\centering
\includegraphics[width=1.0\textwidth]{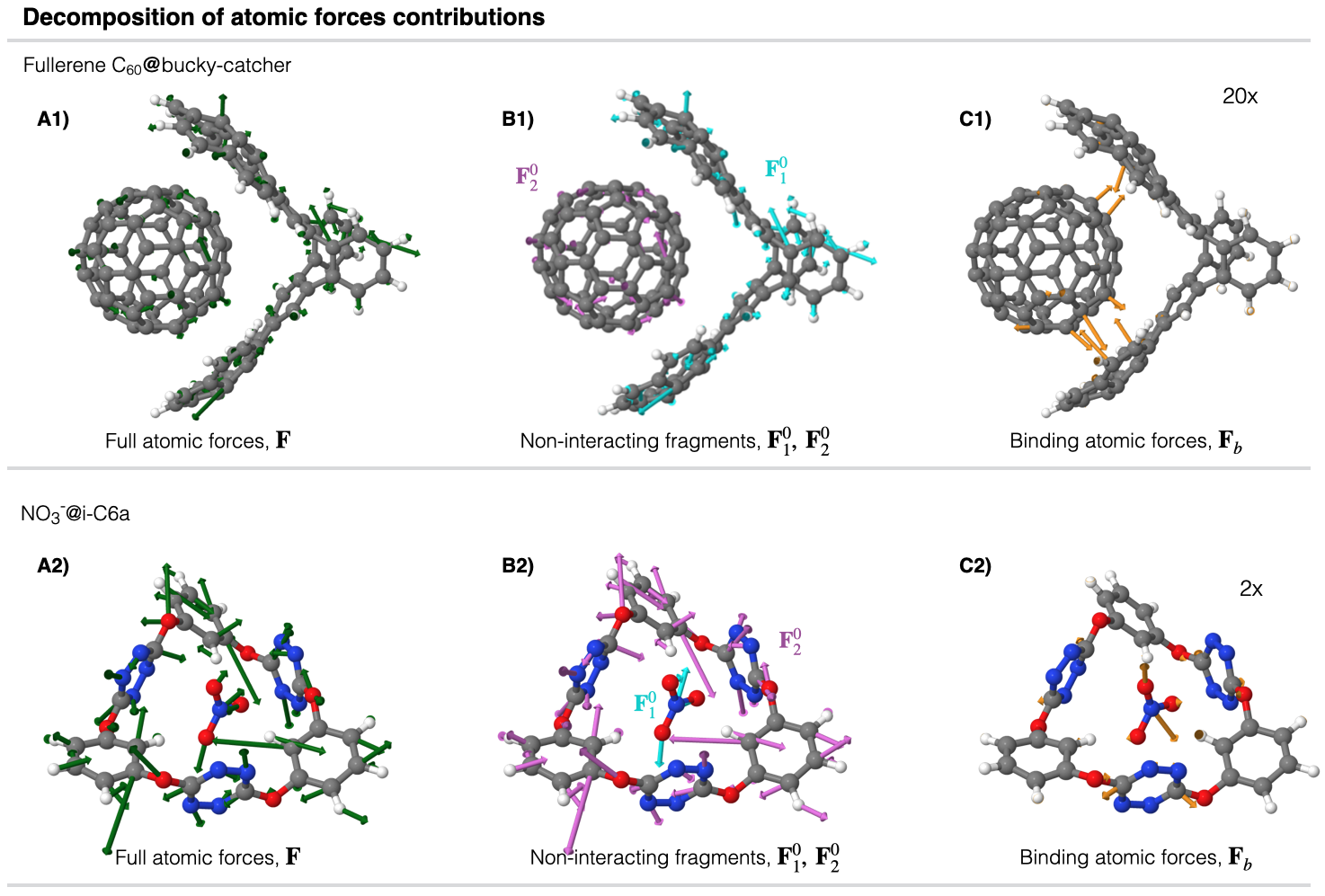}
\caption{Decomposition of atomic force contributions for two representative host-guest systems fullerene C$_{60}$@buckycatcher (top panel/\textbf{1}) and NO$_3^-$@i-corona[6]arene (bottom panel/\textbf{2}). Column A displays full atomic forces $\mathbf{F}(\mathbf{x}_1,\mathbf{x}_2)=(\mathbf{F}^0_1(\mathbf{x}_1),\mathbf{F}^0_2(\mathbf{x}_2))+\mathbf{F}_b(\mathbf{x}_1,\mathbf{x}_2)$, and column B shows the atomic forces for isolated fragments, i.e. $\mathbf{F}^0_1(\mathbf{x}_1)$ was computed without the fragment $\mathbf{x}_2$ and vise-versa. Column C shows the binding atomic forces $\mathbf{F}_b$ between the two fragments. Given the considerable differences in magnitude between covalent and binding forces, an amplification of 20$\times$ and 2$\times$ was applied to $\mathbf{F}_b$ for C$_{60}$@buckycatcher and NO$_3^-$@i-corona[6]arene, respectively.}
\label{fig:Fig_visualFroces}
\end{figure*}

Predictive molecular simulation underpins discovery across chemistry and materials science, driving our understanding of catalysts, supramolecular assemblies, ionic conductors, and complex materials.\cite{ChemRev_Deringer2021,2021_Musil_ChemRev,2025_MLFF-chap_Wiley}
Fundamental to this endeavor is the construction of faithful potential energy surfaces (PESs) that deliver \emph{both} accurate energies and forces across the relevant configurational space at an affordable computational cost, enabling predictive thermodynamics, spectroscopy, kinetics, and dynamics.\cite{2021_Unke_ChemRev}
Within this landscape, a central challenge is the accurate representation of noncovalent (binding) interactions, whose subtle energetics and long-range character strongly influence PES fidelity and emergent behavior.
The realistic description of these forces is particularly consequential, as these terms often dictate emergent behavior in molecular and materials simulations.
Their complexity spans weak yet cumulative van der Waals (vdW) dispersion\cite{2005_book-vdW_parsegian,2016_Science_tkatchenko,2019_Martin_SciAdv,2025_PowerMBD_Tkat}, typically one to two orders of magnitude weaker (in force) than covalent terms, to local, directional motifs—hydrogen, halogen, and ionic bonds; ion$\cdots\pi$, $\pi\cdots\pi$, and polar$\cdots\pi$ interactions\cite{2023_Unconventional-binding_ACSomega,2025_cation-pi_ChemRev}—that can approach covalent force scales (Fig.~\ref{fig:Fig_visualFroces}).

Over the past two decades, machine-learned force fields (MLFFs) have transformed this landscape. Kernel methods,\cite{2010_Bartok_PRL,2015_Bartok2-GAP_IJQC,2017_gdml,2022_bigdml,2024_FFLUX} message-passing \emph{invariant}\cite{2017_SchNetNIPS,2018_SchNet} and \emph{equivariant} neural networks (MPNNs)\cite{2021_PaiNN,2021_SpookyNet,2022_NequiP,2022_MACE,2023_Allegro,2024_SO3krates} now reach near \textit{ab initio} accuracy at dramatically reduced cost. Yet routes to accuracy differ: scalable MPNNs typically adopt locality via a cutoff and then reintroduce nonlocal effects through attention or analytic corrections,\cite{2021_HDNN4,2021_SpookyNet,2021_SOAP-vdW,2024_SO3krates,2025_SO3LR} which may not fully recover the many-body character of dispersion across configuration space. By contrast, \emph{global} MLFFs—such as symmetric gradient-domain ML (sGDML)\cite{2017_gdml,2018_sgdml,2023_gdml_numeric}—learn forces directly and, in principle, can represent long-range correlations end-to-end, trading higher parameter counts for unified expressivity.

\begin{figure}[ht]
\centering
\includegraphics[width=0.98\columnwidth]{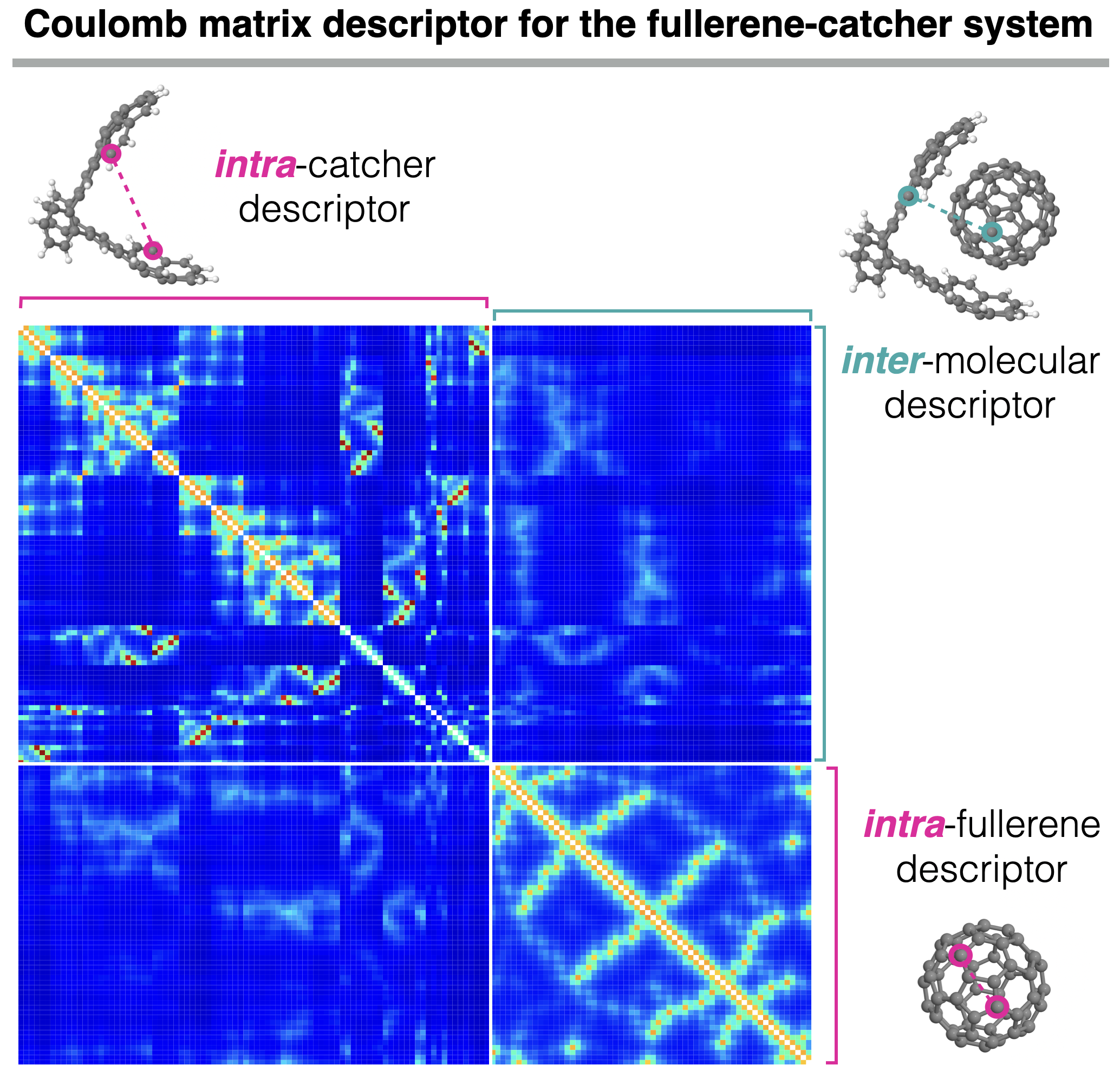}
\caption{Coulomb-matrix (CM) descriptor for the buckyball–catcher complex.
Diagonal blocks encode \textit{intra}-fragment terms; the off-diagonal block encodes \textit{inter}-fragment terms.
Because ${\mathcal D}_{ij}\sim 1/\lVert\mathbf r_i-\mathbf r_j\rVert$, short covalent distances dominate the representation (bright red), while the interfragment part exhibits lower, more homogeneous intensities, which can down-weight nonbonded interactions in joint training.}
\label{fig:Fig_CM}
\end{figure}

In practice, learning \emph{noncovalent} interactions jointly with covalent forces remains difficult because these contributions operate at distinct length scales and force magnitudes (Fig.~\ref{fig:Fig_visualFroces}). When mixed into a single objective, short-range covalent interactions dominate, biasing force vectors and inflating fragment-wise errors. As we will demonstrate, this is particularly acute for global descriptors such as the Coulomb matrix (CM, Fig.~\ref{fig:Fig_CM}) compared under a Euclidean (Frobenius) metric: while the CM excels for single molecules,\cite{2018_sgdml,2021_GDML_Vassilev,2024_GDML_Pandey,2023_GDML_Kazuumi,2019_sgdml_Sauceda_JCP,2020_sgdml_Sauceda_JCP} in multi-fragment systems it overemphasizes intermolecular features, even though in vdW-dominated cases $\sim$95\% of the total force arises from intramolecular (covalent) contributions.

In this work, we address the strength/length-scale mismatch in global models by introducing a $\Delta$-learning formulation within sGDML (\emph{$\Delta$-sGDML}).\cite{2017_gdml,2018_sgdml,2023_gdml_numeric}
Rather than learning all interactions jointly, we \emph{explicitly decouple} the problem: fragment-specific models learn intramolecular physics, and a dedicated \emph{binding} model learns the noncovalent interaction. Energies and forces are then composed additively at inference, preserving global consistency while avoiding length-scale mixing during training. This design retains the long-range expressivity of global models and turns the CM’s amplified intermolecular entries into a useful inductive bias for the binding force field.

We validate the approach on diverse systems--benzene dimers, host–guest complexes (fullerene C$_{60}$@buckycatcher, NO$_3^-$@i-corona[6]arene), benzene-water, and benzene–Na$^+$--and observe consistent fragment-level force improvements over full sGDML models, with \emph{reductions up to \textbf{70–75\%}} while improving or maintaining energy accuracy. Dynamically, the $\Delta$-model yields physically reliable free-energy surfaces and stable trajectories, whereas the single global model overestimates binding forces and generates unstable simulations. 
Overall, decoupling intra- and interfragment learning rebalances the problem and enables global MLFFs to be simultaneously accurate, dynamically robust, and transferable across noncovalent regimes.

\section{Results and Discussion}
%

\subsection{The length-scales problem in learning molecular interactions}\label{TheoryA}

%
\begin{figure*}[htp]
\includegraphics[width=\textwidth]{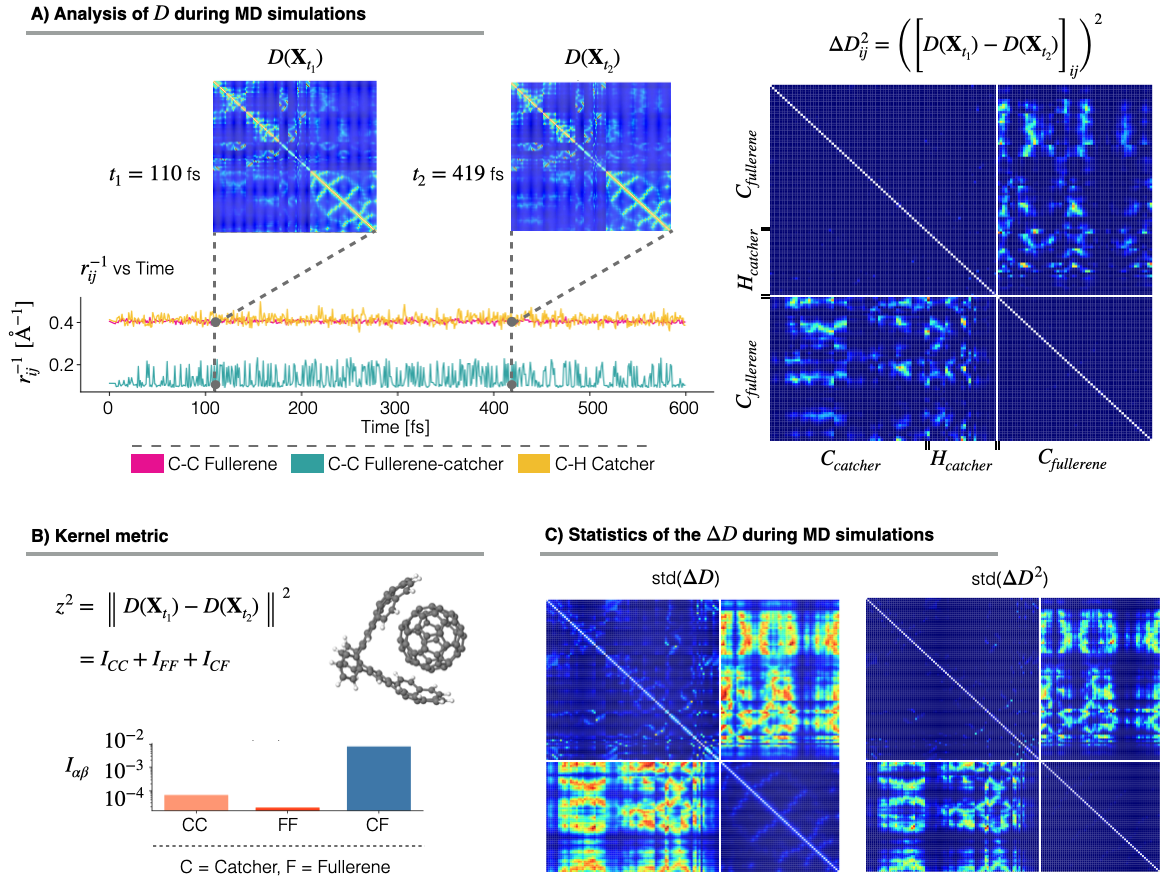}
\caption{Analysis of how the Coulomb–matrix (CM) descriptor biases the similarity metric towards intermolecular features.
\textbf{A)} Time series of three CM entries $[1/r]_{ij}$ along a 300~K \textit{ab initio} MD trajectory of C$_{60}$@buckycatcher: a first-neighbor C–C distance within C$_{60}$ (pink), a first-neighbor C–H within the buckycatcher (yellow), and an interfragment C(catcher)–C(fullerene) pair (cyan). From this trajectory we select two configurations, $D(\mathbf X_{t_1})$ and $D(\mathbf X_{t_2})$; the right panel shows $\Delta D^2=[D(\mathbf X_{t_1})-D(\mathbf X_{t_2})]^2$, where the interfragment block dominates.
\textbf{B)} Decomposition of the squared Frobenius distance used in kernel methods,
$z^2=\lVert\Delta\mathcal D\rVert_F^2 = I_{\mathrm{FF}}+I_{\mathrm{CC}}+I_{\mathrm{CF}}$,
into intrafragment (FF, CC) and interfragment (CF) contributions for the right panel in A (bar plot).
\textbf{C)} Statistics over many random pairs from the same MD: per-entry standard deviation of $\Delta D$ (left) and of $\Delta D^2$ (right). Interfragment entries exhibit the largest fluctuations, i.e., $\sigma_{\mathrm{inter}}\!\sim\!3\times\sigma_{\mathrm{intra}}$. Note the different scales between panels.
}
\label{fig:CM_MD-analysis}
\end{figure*}
Molecular interactions span dissimilar length scales: short-range (e.g. strongly anisotropic covalent and exchange–repulsion terms), intermediate-range induction and specific directional motifs (e.g. hydrogen/halogen bonding, ion$\cdots\pi$), and long-range (e.g. electrostatics and dispersion).
Therefore, jointly reconstructing such a variety of interactions from the data using only atomic coordinates is a primary challenge.
%
%
Global MLFFs, in principle, learn the many-body forces (i.e. quantum mechanical observables: $\mathbf{F}=-\langle\Psi|\partial\mathcal{H}/\partial\mathbf{x}|\Psi\rangle$) end-to-end from the data. Meaning that they could learn all these interactions at once.
Success then hinges on whether the \emph{descriptor} provides a faithful multiscale representation.
Figure~\ref{fig:Fig_CM} shows the Coulomb–matrix (CM) descriptor for an equilibrium C$_{60}$@buckycatcher supramolecule configuration. Intrafragment blocks (diagonal) have larger raw values than the interfragment block, as expected, because noncovalent forces are typically one or two orders of magnitude weaker than covalent forces. However, when comparing configurations in kernel methods, the similarity metric is the Euclidean (Frobenius) distance,
$z_{ij}=\lVert \mathcal D(\mathbf X_i)-\mathcal D(\mathbf X_j)\rVert_F$,
and along an MD trajectory the situation inverts: the interfragment entries dominate the $z_{ij}$ metric, while intrafragment terms dominate the raw descriptor.
This is illustrated in Fig.~\ref{fig:CM_MD-analysis}A. The intrafragment entries $D_{C_{\mathrm C}H_{\mathrm C}}$ and $D_{C_{\mathrm F}C_{\mathrm F}}$ oscillate around tight means, whereas the interfragment entry $D_{C_{\mathrm C}C_{\mathrm F}}$ has a lower mean but markedly higher variance. Consequently, the interfragment block dominates the squared difference $\Delta D^2=[D(\mathbf X_{t_1})-D(\mathbf X_{t_2})]^2$ (right panel). We make this explicit in Fig.~\ref{fig:CM_MD-analysis}B by decomposing the squared distance
\[
z^2 \;=\; \lVert\Delta \mathcal D\rVert_F^2 \;=\; I_{\mathrm{FF}} + I_{\mathrm{CC}} + I_{\mathrm{CF}},
\]
where $I_{\mathrm{FF}}$ and $I_{\mathrm{CC}}$ collect intrafragment contributions and $I_{\mathrm{CF}}$ the interfragment ones: $I_{\mathrm{CF}}$ dominates by orders of magnitude. Aggregating over many random pairs from the same MD (Fig.~\ref{fig:CM_MD-analysis}C) confirms that interfragment entries exhibit the largest fluctuations, with $\sigma_{\mathrm{inter}}\sim 3\times\sigma_{\mathrm{intra}}$.
Thus, although the CM is well behaved for \emph{single} molecules, in multifragment systems coupling it with a Frobenius metric skews similarity toward the off-diagonal (interfragment) block, amplifying intermolecular changes while underweighting intrafragment ones. Concretely, even though intramolecular forces are $\sim 10$–$100\times$ larger than noncovalent forces, the kernel metric emphasizes nonbonded molecular features. Therefore, the mismatch between molecular representation and atomic force strengths imposes trade-offs when training a single global model which bias learning and degrades performance.

To address this, we introduce a $\Delta$-learning framework that trains separate submodels for distinct interaction types and length scales (intrafragment vs.~binding), thereby mitigating cross-scale mixing in the metric and improving accuracy.

\begin{figure}[ht]
\centering
\includegraphics[width=0.98\columnwidth]{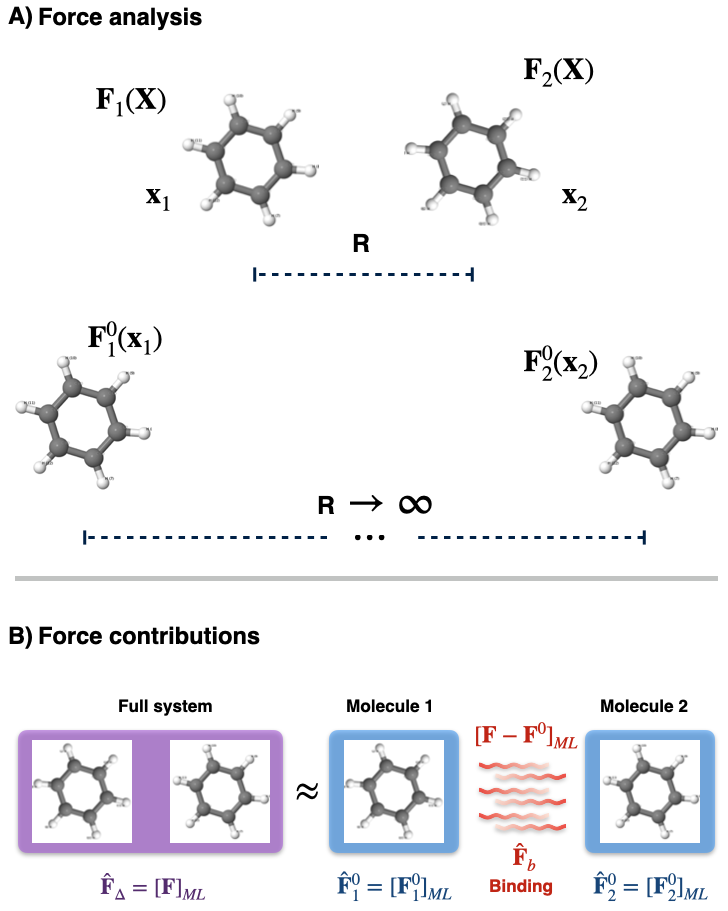}
\caption{\textbf{A)} Decoupling intrafragment force fields by increasing the interfragment separation $R\to\infty$: the forces on fragment~1 (2) become independent of the coordinates of fragment~2 (1). \textbf{B)} Definition of the $\Delta$-learned force field ($\Delta$MLFF) as the sum of noninteracting intrafragment forces and a binding (interfragment) term.}
\label{fig:Fig_DeltaForces}
\end{figure}

\subsection{Range-separated molecular interactions: The binding force field}

Consider two noncovalently bound fragments with coordinates $\mathbf x_1$ and $\mathbf x_2$ separated by an effective distance $R$, and denote the full system by $\mathbf X=(\mathbf x_1,\mathbf x_2)$ (Fig.~\ref{fig:Fig_DeltaForces}A). The total atomic forces are $\mathbf F(\mathbf X)=\big(\mathbf F_1(\mathbf X),\,\mathbf F_2(\mathbf X)\big)$.
In the noninteracting limit $R\to\infty$ the fragments no longer exert forces on each other, defining the \emph{noninteracting} force field
$\mathbf F^{0}(\mathbf X)\equiv \lim_{R\to\infty}\mathbf F(\mathbf X)=\big(\mathbf F^{0}_1(\mathbf x_1),\,\mathbf F^{0}_2(\mathbf x_2)\big)$.
We then separate the force-field $\mathbf{F}(\mathbf{X})$ as,

\begin{equation}
\label{eq:interMolFF}
\mathbf F(\mathbf X)=\mathbf F^{0}(\mathbf X)+\mathbf F_b(\mathbf X),
\end{equation}

\noindent where $\mathbf F_b$ is the \emph{binding} force field that contains all intermolecular contributions (electrostatics, induction, dispersion, etc.). %
Similarly, the \textit{binding energy} is defined as $E_{b}(\mathbf{X})=E(\mathbf{X})-E_1^{0}(\mathbf{x}_1)-E_2^{0}(\mathbf{x}_2)$, where $E$ is the total energy and $E_j^{0}$ is the molecular energy of the isolated fragment $j$. Figure~\ref{fig:Fig_DeltaForces}B summarizes this separation.

Figure~\ref{fig:Fig_visualFroces} shows this decomposition for the atomic forces of two representative molecular systems. The first one is the fullerene C$_{60}$@buckycatcher with binding forces mainly due to dispersion interactions, here intermolecular forces $\mathbf{F}_{b}$ were scaled by 20$\times$ to comparable to intramolecular forces. The second molecule is NO$^{-}_{3}$@i-corona[6]arene, where the fragments interact stronger due to a richer dynamical mix of binding forces (e.g. anion-$\pi$ and hydrogen-bond)~\cite{2020_NO3ic6a}.

\subsection{$\Delta$-learning Framework} \label{deltaML_frame}
$\Delta$-learning ($\Delta$ML) has proven broadly useful mainly for retaining quantum-mechanical accuracy at substantially lower cost by learning corrections to a cheaper baseline. It has been used to extend DFT-level accuracy to larger systems and thermodynamic/dynamic properties (e.g., liquid Ar with $\Delta$-NetFF) \cite{Delta-Net_Pattnaik2020}, improve spectroscopic predictions such as Raman spectra with less training data than direct ML \cite{Delta_Grumet2024}, and 
construct coupled-cluster–quality PESs for molecules including ethanol, acetylacetone, hydronium, and tropolone \cite{Delta_Nandi2022,Delta_Qu2021,NANDI2024100036,Delta_Bowman2023,Delta_Nandi2023,Delta_Lilienfeld2015}. Hierarchical schemes combining multiple $\Delta$ corrections further reduce cost while maintaining accuracy on reactive and complex PESs \cite{Dral2020}. The approach also generalizes to diverse properties, for example, vertical ionization potentials, redox and absorption energies, protein–ligand scoring, and even phase diagrams of correlated bosons~\cite{Delta_Maier2023,Delta_Chen2023,Delta_Yang2022,delta_lin2024}. Additionally, $\Delta$ML  has been used in multi-fidelity strategies to integrate information across levels of theory to accelerate materials discovery and property optimization \cite{Multifidelity2019}.

Here we apply this idea to the reconstruction of a global interactions via a $\Delta-$\textit{learned force-field} ($\Delta$MLFF) $\hat{\mathbf F}_{\Delta}$.
Unsing Eq.~\eqref{eq:interMolFF}, we approximate our global model $\hat{\mathbf{F}}_{\text{full}}$
by learning two different models for the noninteracting fragments and for the binding term:

\begin{equation}
\label{eq:DeltaMLFF}
\hat{\mathbf F}_{\mathrm{full}} \;=\; [\mathbf F]_{\mathrm{ML}}
\;\approx\; \underbrace{[\mathbf F^{0}]_{\mathrm{ML}}}_{\hat{\mathbf F}^{0}}
\;+\;
\underbrace{[\mathbf F-\mathbf F^{0}]_{\mathrm{ML}}}_{\hat{\mathbf F}_b}
\;\equiv\; \hat{\mathbf F}_{\Delta}.
\end{equation}

Here $\hat{\mathbf F}_b$ has the same dimensionality as the original force field, while $\hat{\mathbf F}^{0}=\big(\hat{\mathbf F}^{0}_1,\,\hat{\mathbf F}^{0}_2\big)$ is learned on two smaller, per-fragment datasets.
This approximation is naturally occurring in physics, either mathematically or by a hierarchy of interactions.
In effect, we replace a single joint problem with two specialized interaction–length-scale problems (intra and inter), which improves conditioning and avoids the descriptor–metric bias identified in Sec.~\ref{TheoryA}. 
Related energy decompositions have been used in the study of diatomic molecular scattering and lattice energy calculations for molecular
crystals~\cite{2023_DeltaML_Liu,2022_DeltaGAP}.

\begin{figure*}[ht]
\centering
\includegraphics[width=0.90\textwidth]{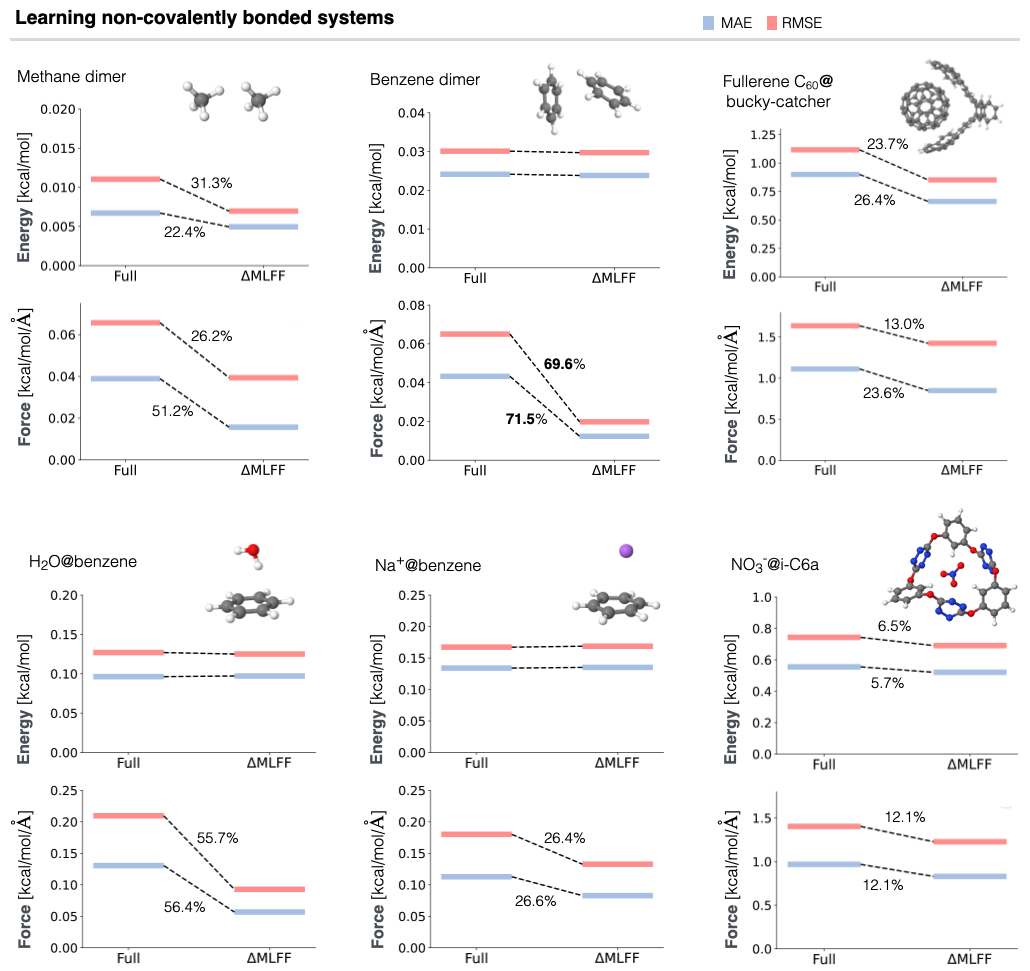}
\caption{Global performance comparison between the composite $\Delta$MLFF and a single full sGDML model.
For each system, the improvement of the $\Delta$MLFF over the full model is reported using the MAE (blue) and RMSE (red) metrics for total energies and atomic forces. Units: kcal\,mol$^{-1}$ (energies) and kcal\,mol$^{-1}$\,\AA$^{-1}$ (atomic forces). Training set sizes (per system): methane dimer (200), benzene dimer (1000), H$_2$O@benzene (1000), Na$^+$@benzene (1000), NO$_3^-$@i-C6a (800), and C$_{60}$@buckycatcher (100).}
\label{fig:Fig_train}
\end{figure*}

In our setting, this interaction length–scales decoupling turns a drawback of the Coulomb matrix descriptor (i.e. the over-representation of intermolecular features in the Frobenius metric) into a useful inductive bias for the learning the binding force field, while retaining the strong per-fragment performance of global models~\cite{2018_sgdml}.
%
To explore the applications of this approach, we use the sGDML framework~\cite{2017_gdml,2018_sgdml,2023_gdml_numeric}, which, combined with the CM descriptor, has proven robust across sizes and fluxional complexity~\cite{2022_bigdml,2021_GDML_Vassilev,2023_GDML_Kazuumi,2024_GDML_Pandey}.
We devise a composite $\Delta$-sGDML model based on Eq.~\ref {eq:DeltaMLFF} to address the length-scale representability problem (Sec.~\ref{TheoryA}) while simulteneusly capturing the global many-body nature of intermolecular interactions. 
As summarized in Fig.~\ref{fig:Fig_train}, the $\Delta$ scheme consistently improves force accuracy across systems with diverse binding physics. Methodological details appear in Sec.~\ref{Methodology}.

\begin{figure}[ht]
\centering
\includegraphics[width=0.98\columnwidth]{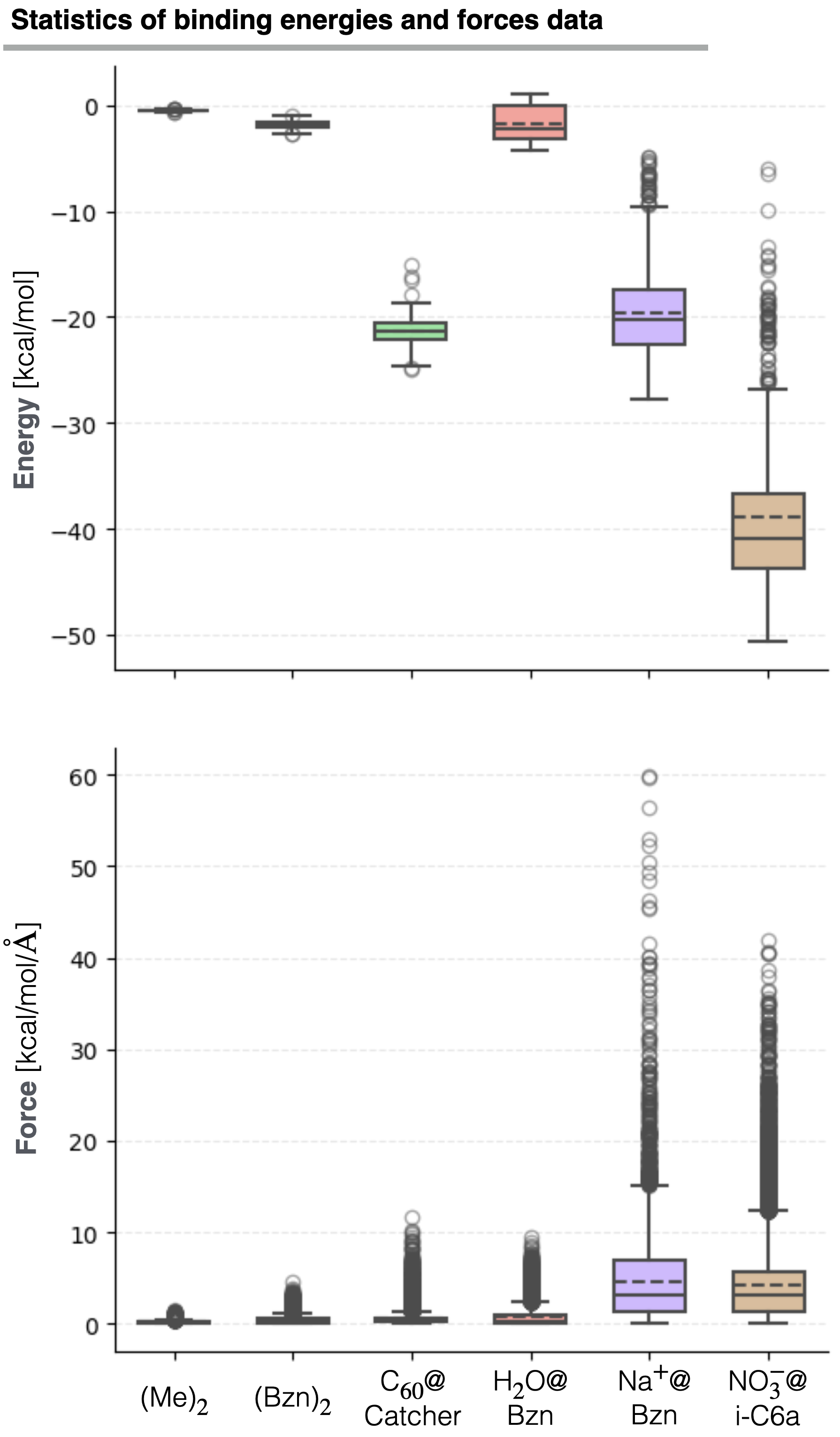}
\caption{Summary statistics of binding energies (top) and binding-force magnitudes (bottom) across datasets. Boxplots show mean (dashed line), median (solid line), and interquartile range (box). Abbreviations: Me = methane; Bzn = benzene; Catcher = buckycatcher; i-C6a = i-corona[6]arene.}
\label{fig:Fig_binding-stats}
\end{figure}
%

\subsection{Datasets}\label{sec:datasets}

We evaluate the $\Delta$MLFF approach within the GDML framework~\cite{2017_gdml,2018_sgdml} on representative noncovalent systems: methane and benzene dimers~\cite{S22}, benzene interacting with water~\cite{S22,S22Brev} and with a sodium cation (Na$^+$)~\cite{cation-pi_1996,cation-pi_JACS1996,cation-pi_ChemRev2013}, and two host–guest complexes, NO$_3^-$@i-corona[6]arene~\cite{2020_NO3ic6a} and C$_{60}$@buckycatcher~\cite{C60-buckycatch,S12L,vdW-S12L}. These systems span diverse noncovalent motifs—van der Waals dispersion, anion/cation–$\pi$, hydrogen bonding, X–H$\cdots\pi$, and $\pi$–stacking~\cite{2024_Noncovalent_interACSomega}.
Figure~\ref{fig:Fig_binding-stats} summarizes, for all six systems, the distributions of binding energies (top) and binding-force magnitudes (bottom). As expected for dispersion-dominated cases (methane dimer, benzene dimer, C$_{60}$@buckycatcher), the binding-energy and force distributions are relatively compact. In contrast, systems with stronger and more directional interactions (X–H$\cdots\pi$, cation–$\pi$), or mixed motifs as in NO$_3^-$@i-corona[6]arene~\cite{2020_NO3ic6a}, exhibit considerably broader spreads.
A noteworthy case is H$_2$O@benzene, where binding energies frequently take \emph{positive} values because the MD sampling includes short-range, repulsive configurations (e.g. Pauli/steric repulsion or unfavorable orientation at short distances, where the water dipole and the benzene quadrupole can be repulsive).

Overall, the datasets cover a broad range of interaction strengths and configurational complexity—well suited to stress-test our methodology. All configurations were generated via DFT-based molecular dynamics. Further details (levels of theory, thermostats, time steps) are provided in Sec.~\ref{Methodology}.

\begin{figure*}[ht]
\centering
\includegraphics[width=1.0\textwidth]{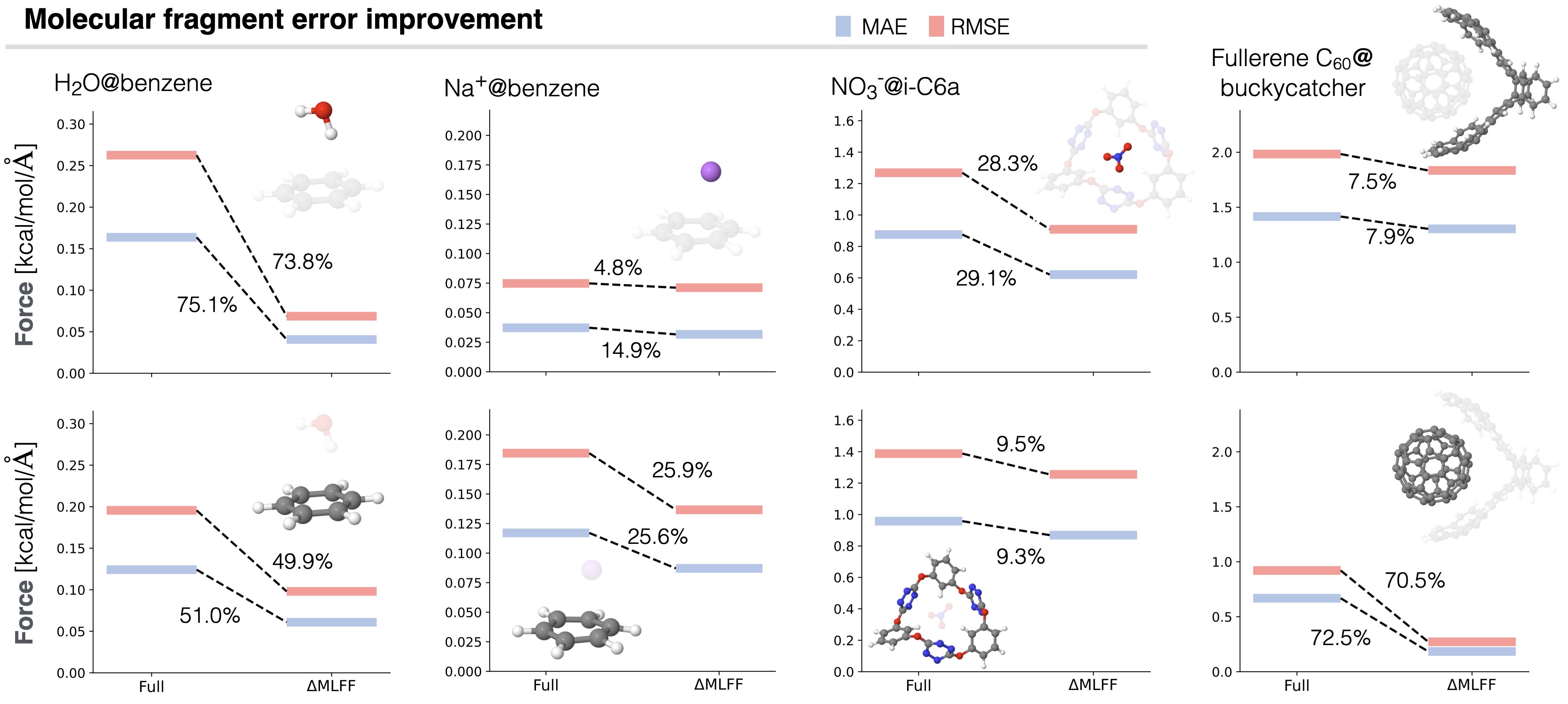}
\caption{Comparison of fragment-resolved atomic-force errors between the $\Delta$-learned force field and the full model. For each system (columns), the improvement of the $\Delta$MLFF over the full model is reported per fragment (top and bottom panels). Units are kcal mol$^{-1}$ \AA$^{-1}$.}
\label{fig:Fig_errFragment}
\end{figure*}

\subsection{Training $\Delta$MLFF} \label{training}

To assess the benefit of $\Delta$-learning, we trained two models per system: (i) a single full sGDML model, $\hat{\mathbf F}_{\mathrm{full}}$, on the entire supramolecule, and (ii) a composite $\Delta$-sGDML model, $\hat{\mathbf F}_{\Delta}$, assembled from intrafragment and binding submodels (Sec.~\ref{deltaML_frame}). 
Global test errors are summarized in Fig.~\ref{fig:Fig_train} (see Table~S1 for exact values).

Decoupling intramolecular and binding interactions consistently improves atomic-force accuracy, with gains up to \textbf{71.5\%} in MAE. 
Dispersion-dominated systems—methane dimer, benzene dimer, and C$_{60}$@buckycatcher—show the largest improvements. 
For methane dimer, force MAE (RMSE) decreases by 51.2\% (26.2\%), and energy MAE (RMSE) by 22.4\% (31.3\%). 
For benzene dimer, force MAE (RMSE) improves by \textbf{71.5\%} (69.6\%). 
These results highlight the value of separating intra- from intermolecular interactions and directly address the descriptor–metric bias identified in Sec.~\ref{TheoryA}.

For C$_{60}$@buckycatcher, $\Delta$-sGDML yields consistent gains of $\sim$24\% overall, with a 13\% improvement in force RMSE. 
The complexity of modeling the buckycatcher in higher-energy regions of its configurational space likely contributes to this RMSE error; we return to this in Sec.~\ref{per-frag-err} via fragment-resolved errors.
By contrast, NO$_3^-$@i-corona[6]arene shows smaller but clear gains--$\sim$6\% in energies and $\sim$13\% in forces--consistent with stronger, directional interactions and broader binding-forces (see Fig.~\ref{fig:Fig_binding-stats}) within the macrocyclic cavity substantially reshape the host’ structure and dynamics relative to the isolated molecule.

Benzene-containing systems further illustrate the trend: benzene dimer and H$_2$O@benzene achieve large force-error reductions (71.5\%/69.6\% and 56.0\%/55.7\% for MAE/RMSE, respectively), whereas Na$^+$@benzene improves by $\sim$26\%. 
In these three systems, energy gains are statistically negligible ($\sim\!\pm 1\%$), consistent with prior observations that energy accuracy saturates early with training size for benzene datasets, while force accuracy continues to improve with more data~\cite{2018_sgdml,2018_SchNet}.

Overall, Fig.~\ref{fig:Fig_train} demonstrates that $\Delta$-learning substantially strengthens kernel models for systems that mix strong covalent with much weaker noncovalent interactions (e.g., dispersion), mitigating the mismatch between the CM–Frobenius metric (which overemphasizes intermolecular features) and the labels’ dominant covalent contributions. 
Although global metrics might suggest the benefits are largest for weak binding, a fragment-wise analysis (Sec.~\ref{per-frag-err}) reveals a more informative picture and concrete levers for future systematic improvement.

\begin{figure*}[ht]
\centering
\includegraphics[width=1.0\textwidth]{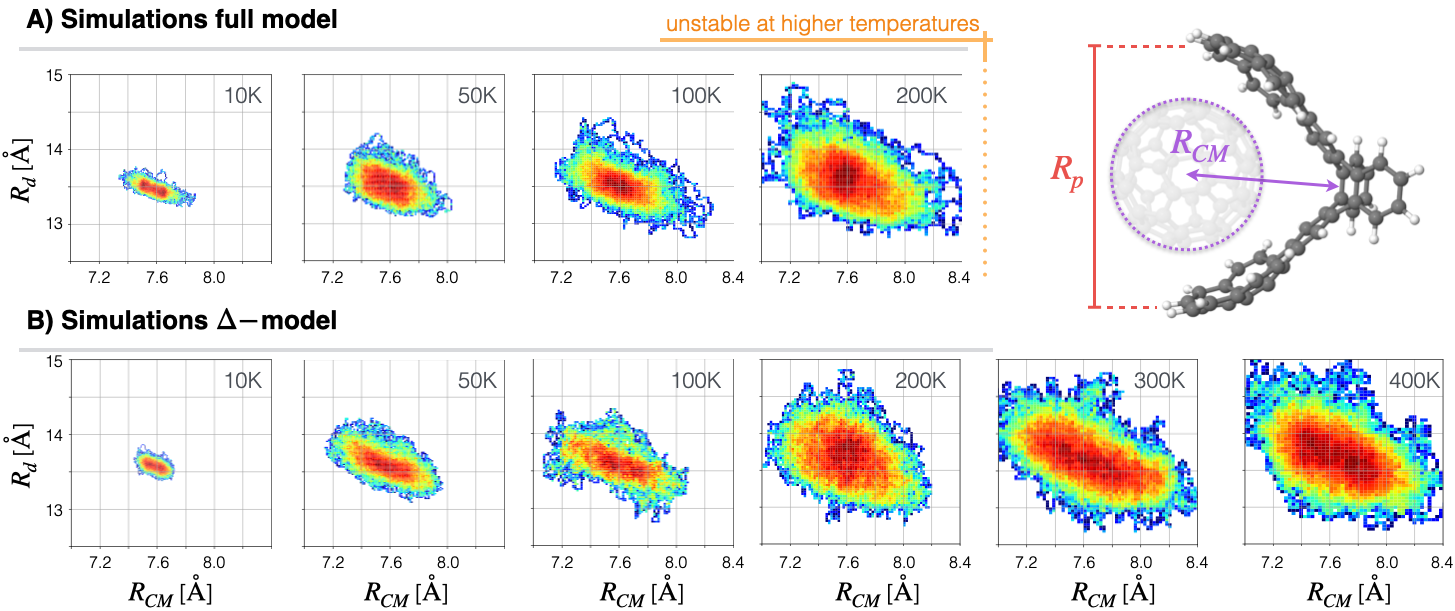}
\caption{Molecular dynamics simulations of the C$_{60}$@buckycatcher system using the \textbf{A}) full sGDML and \textbf{B}) the $\Delta$-learned force fields for different temperatures. Panels show the free energy surfaces projected to the variables: $R_p$, measuring the opening buckycatcher's \textit{clamps} and $R_{CM}$ to quantify the distance between the C$_{60}$ center of mass to the buckycatcher.}
\label{fig:Fig_MDC60BC}
\end{figure*}

\subsection{Performance improvement per molecular fragment}\label{per-frag-err}

Global metrics (Fig.~\ref{fig:Fig_train}) quantify overall gains, but for noncovalently bound systems it is equally important to ask how much \emph{each} fragment benefits. Figure~\ref{fig:Fig_errFragment} reports fragment-resolved force errors, comparing the composite $\Delta$MLFF against a single full model.

Two clear trends emerge. First, $\Delta$MLFF consistently reduces force errors for \emph{both} fragments across all systems, but the magnitude of the improvement is asymmetric and correlates with molecular complexity and size: smaller or less complex partners (H$_2$O, NO$_3^-$, C$_{60}$) gain the most—up to \textbf{75\%}—whereas larger hosts (buckycatcher, i-corona[6]arene) improve modestly (typically $<\!10\%$). This \emph{fragment-resolved error asymmetry} likely stems from single-model training, where the loss is dominated by the larger fragment (more atoms, broader force variance), leading to systematic underfitting of the smaller partner. By decoupling intra- and interfragment learning, $\Delta$MLFF counteracts this imbalance and yields more fragment-faithful forces.
Second, when benzene is paired with different fragments (Na$^+$, H$_2$O, benzene), the benzene force error follows the strength of the binding motif. Stronger binding forces induce larger deformations and shifts sampling toward regions of higher PES curvature, making forces harder to learn. Consistent with this picture, the benzene force MAE increases in the order $\text{bzn}\cdots\text{bzn} \;<\; \mathrm{H_2O}\cdots\text{bzn} \;<\; \mathrm{Na^+}\cdots\text{bzn}$,
with MAE values \(0.0123\), \(0.0481\), and \(0.0841~\mathrm{kcal\,mol^{-1}\,\AA^{-1}}\), respectively.
These trends align with the binding-energy/force statistics in Fig.~\ref{fig:Fig_binding-stats}.

Taken together, the fragment-wise analysis confirms that (i) single, global training tends to favor the larger fragment under CM+Frobenius metrics, and (ii) $\Delta$-learning systematically mitigates this bias, improving force fidelity and, consequently, more reliable force fields. In the next section, we perform molecular dynamics to test the stability of the simulations for both models.

\subsection{Dynamical stability of molecular simulations: Full vs. $\Delta$-model}

Accurate forces must yield stable and physically plausible dynamics across temperature. 
We therefore ran classical MD for C$_{60}$@buckycatcher using both the single full sGDML model and the composite $\Delta$-sGDML model over a broad range of temperatures (10--400~K), and analyzed (i) two–dimensional free-energy surfaces (FES) projected onto the clamp-opening coordinate $R_p$ and the center-of-mass separation $R_{\mathrm{CM}}$, and (ii) the mean-squared angular displacement (MSAD) of C$_{60}$ as a proxy for rotational diffusion (Fig.~\ref{fig:Fig_MDC60BC}).

\paragraph*{Overbinding and hindered rotation in the full model.}
At 300~K, the MSAD of C$_{60}$ is substantially lower with the full model ($1.39\times 10^{-5}$) than with the $\Delta$-model ($3.41\times 10^{-5}$), indicating suppressed rotational mobility. 
Consistently, the FES obtained with the full model is more localized in $(R_p,R_{\mathrm{CM}})$, generating deeper basins that confine the guest (Fig.~\ref{fig:Fig_MDC60BC}A). 
Both observations are consistent with an \emph{overestimation of interfragment binding forces} by the single global model, in line with the descriptor–metric bias identified in Sec.~\ref{TheoryA}.

\paragraph*{Thermal robustness and sampling.}
A more striking difference appears upon increasing the simulation temperature: trajectories driven by the full model lose stability above $\sim 200$~K, whereas the $\Delta$MLFF yields stable dynamics throughout the entire 10–400~K range (Fig.~\ref{fig:Fig_MDC60BC}B). 
The $\Delta$-model’s dynamics remain smooth and physically balanced across temperatures, enabling rotational diffusion of C$_{60}$ and clamp breathing of the host, as expected for a dispersion-dominated host–guest system.

These results provide a dynamical stress test that complements the static error metrics: the single global model’s tendency to overweight interfragment variance (under CM+Frobenius) manifests as overestimation of the attractive binding landscapes that hinder rotation and precipitate instabilities at elevated temperatures.
By decoupling intra- and interfragment learning, the $\Delta$MLFF restores the proper scale of noncovalent forces, improves rotational mobility, and ensures stable MD over a wider thermal window—critical for reliable sampling and thermodynamic.

\section{Conclusions}

We have introduced a $\Delta$-learning formulation within the sGDML framework (\emph{$\Delta$-sGDML}) that \emph{explicitly decouples} intra- and intermolecular interactions by composing fragment-specific models with a dedicated global binding force field [Eq.~(\ref{eq:DeltaMLFF})]. This design directly addresses the contradictory intrafragment covalent nature of the data and the CM–Frobenius metric bias towards intermolecular features in single global models, thereby improving conditioning and restoring the proper balance between covalent and noncovalent forces.

Across dispersion-dominated dimers (methane, benzene) and heterogeneous host–guest and ion–$\pi$ systems (C$_{60}$@buckycatcher, NO$_3^-$@i-corona[6]arene, H$_2$O@benzene, Na$^+$@benzene), $\Delta$-sGDML consistently reduces atomic-force errors relative to a single full model—often by tens of percent and up to \textbf{75\%} fragment-wise—while maintaining energy accuracy. Fragment-resolved analysis reveals a systematic asymmetry in single-model training (the larger fragment dominates the loss), which $\Delta$-learning mitigates by construction, yielding more \emph{fragment-faithful} forces. Furthermore, molecular dynamnics simulations provide an independent stress test for the models beyond static metrics. For C$_{60}$@buckycatcher, consistent with our initial hipothesis (Sec.~\ref{TheoryA}) the full model overestimates the strenght of the binding forces and hinders C$_{60}$ rotations (lower MSAD) and becomes unstable above $\sim$200~K, whereas the $\Delta$-model produces stable trajectories from 10–400~K. Thus, the static accuracy gains of $\Delta$-sGDML translate into improved sampling robustness and thermodynamic reliability.

Methodologically, the approach offers a practical route to \emph{homogenize per-fragment errors} without sacrificing global consistency or symmetry handling, and it remains fully compatible with alternative descriptors and kernels. Given its modularity, the same recipe can be extended to multifragment assemblies, mixed long-range corrections, and broader benchmarks (e.g., S22~\cite{S22}, S66~\cite{S66}, L7~\cite{L7} for molecular complexes; C21~\cite{C21} and X23~\cite{X23} for molecular solids with hydrogen-bonded and vdW character), facilitating systematic studies of transferability.
The results presented here open a direct avenue for improving our understanding of MLFFs and their inner workings, as well as going beyond kernels and combining $\Delta$-learning with equivariant message-passing models, and active learning that targets binding subspaces with maximal uncertainty. Furthermore, $\Delta$MLFFs open the possibility to transfer fragment force fields to other composite models, and to use them as the basis to construct more general force fields describing molecular crystals or liquids.
We anticipate that $\Delta$-learning force fields will become a standard ingredient for reliable and scalable machine learned models in non-reactive heterogeneous systems, enabling to build on this modular recipe across broader materials and applications.

\section{Methodology} \label{Methodology}
\subsection*{Data generation and DFT calculations}
The databases, except the fullerene C$_{60}$@buckycatcher, were generated using molecular dynamics simulations (MD) with an integration step of 1.0 fs and the NVT thermostat using the DFT level of theory with the Perdew-Burke-Ernzerhof (PBE)~\cite{1996_PBE} exchange-correlation functional with the non-local many body dispersion (MBDnl)~\cite{2020_MBDnl,2012_MBD,2014_MBD} treatment of the van der Waals interaction and the \textit{light} basis set using the FHI-aims~\cite{2019_FHIaims} code.
Due to the nature of the weak binding energies, the databases were generated at MD temperatures in the range 100K (for methane dimer), 150K (for benzene dimer), 300K (for Na$^+$@benzene), 500K (for H2O@benzene and NO$_3^-$@i-corona[6]arene).

In the fullerene C$_{60}$@buckycatcher case, we subsampled a set of molecular configurations from the MD trajectory from the MD22 database~\cite{2023_gdml_numeric}, and then recalculated the energies and forces using a larger basis set (from \textit{light}$\to$\textit{really\_tight}) at the PBE+MBDnl level of theory.
Similarly, for the benzene-dimer system, we subsampled its MD trajectory and then refined the accuracy of the database using the PBE0+MBDnl/\textit{really\_tight} level of theory, as described in Ref.~\cite{2020_sgdml_bookAppl}.

The results displayed in Fig.~\ref{fig:Fig_MDC60BC} were obtained using MD simulations for 500 ps using an integration step of 0.5 fs using the sGDML interface\cite{2019_sGDMLsoftware} to the i-PI software.~\cite{2014_iPI,2018_iPI} 

\subsection*{Database Construction}

\begin{itemize}
    \item Get the dataset $\mathcal{W}=\{\mathbf{X}_i,\mathbf{F}_i, E_i\}_{i=1}^{M}$.
    \item Assuming that the atomic coordinates are sorted in groups for the two molecular components (e.g., first the water molecule and then benzene, $\mathbf{X}_i=\mathbf{x}_i^{(H_2O)}\cup\mathbf{x}_i^{(benzene)}$, split the coordinates for each of the component in the system $\{\mathbf{X}_i=\mathbf{x}_i^{(1)}\cup\mathbf{x}_i^{(2)}\}_{i=1}^{M}$ to construct the new sets of coordinates for the fragments: $\{\mathbf{x}_i^{(1)}\}_{i=1}^{M}$ and $\{\mathbf{x}_i^{(2)}\}_{i=1}^{M}$.
    \item Create (compute) the new databases for each fragment: $$\mathcal{W}^{(1)}=\{\mathbf{x}_i^{(1)}, \mathbf{F}_i^{(1)}, E_i^{(1)}\}_{i=1}^{M}$$ and $$\mathcal{W}^{(2)}=\{\mathbf{x}_i^{(2)}, \mathbf{F}_i^{(2)}, E_i^{(2)}\}_{i=1}^{M}$$
    
    \item From these, assemble the non-interacting base force field data set: $$\mathcal{W}^0=\{\mathbf{x}_i^{(1)}\cup\mathbf{x}_i^{(2)},\mathbf{F}_i^{(1)}\cup\mathbf{F}_i^{(2)}, E_i^{(1)}+E_i^{(2)}\}$$
    $$=\{\mathbf{X}_i, \mathbf{F}_i^{0}, E_i^0\}_{i=1}^{M}$$
    
    \item Create the interaction or \textit{binding} force field dataset: $$\mathcal{W}^{b}=\{\mathbf{X}_i,\mathbf{F}_i-\mathbf{F}_i^{0}, E_i-E_i^{0}\}_{i=1}^{M}$$ $$=\{\mathbf{X}_i,\mathbf{F}_{b,i}, E_{b,i}\}_{i=1}^{M}$$
    
    \item Train the individual models $\hat{f}_{\mathbf{F}}^{(j)}$ for each fragment using the non-interacting databases $\mathcal{W}^{(j)}$.
    
    \item Assemble the non-interacting force field: $$\hat{f}_{\mathbf{F}}^{0}=(\hat{f}_{\mathbf{F}}^{(1)},\mathbf{0})^T+(\mathbf{0},\hat{f}_{\mathbf{F}}^{(2)})^T$$
    $$=(\hat{f}_{\mathbf{F}}^{(1)},\hat{f}_{\mathbf{F}}^{(2)})^T=(\hat{\mathbf{F}}_1^{0},\hat{\mathbf{F}}_2^{0})^T=\hat{\mathbf{F}}^{0}$$
    
    \item Train the binding force field model $\hat{f}_{\mathbf{F}_b}$ using the binding interaction database $\mathcal{W}^{b}$:  $\hat{f}_{\mathbf{F}_b}=[\mathbf{F}_{b}]_{ML}=\hat{\mathbf{F}}_{b}$
    
    \item Construct and test the composite model $\hat{\mathbf{F}}_{\Delta}=\hat{\mathbf{F}}^{0}+\hat{\mathbf{F}}_b$.
    
\end{itemize}

\subsection{Ion-benzene case}
The total energy from a $\Delta$-model is $\hat{E}_{\Delta}=\hat{E}_{1}^0+\hat{E}_{2}^0+\hat{E}_{b}$, where, for ions (e.g. Li$^+$ and Na$^+$), $\hat{E}_{1}^0\equiv E_{ion}$, the atomic energy computed at the same theory level as the rest of the calculations and its value is constant in the model. 
Thereby, and because it has no atomistic internal structure, in Eq.~\ref{eq:DeltaMLFF}, $\hat{\mathbf{F}}^{0}=(\mathbf{0},\hat{\mathbf{F}}^{0}_{benzene})^T$.

\section{ACKNOWLEDGEMENTS}
H.E.S. acknowledges support from CONACYT/SECIHTI-Mexico under Project CF-2023-I-468, DGTIC-UNAM under Project LANCAD-UNAM-DGTIC-419, DGAPA-UNAM PAPIIT No. IA106023 and No. IA105625. H.E.S. acknowledges Carlos Ernesto L\'opez Natar\'en for helping with the high-performance computing infrastructure. L.C.T. and M.L.S. acknowledge DGAPA-UNAM for the scholarship given during the development of the project PAPIIT No. IA106023. We also thank Dr. Claudia Islas-Vargas for her insightful discussions and comments on the article.

\bibliography{references/clean_ref}
%


\end{document}